\RequirePackage{fix-cm}
\documentclass[smallextended]{svjour3}       
\smartqed  
\usepackage{graphicx}
\usepackage{bm}
\usepackage{mathbbol}
\usepackage{braket}
\usepackage{amssymb,amsmath,amsfonts}
\newcommand{\up}{\uparrow}
\newcommand{\down}{\downarrow}

\begin{document}

\title{Asymptotic entanglement in quantum walks from delocalized initial states
}


\author{Alexandre C. Orthey Jr.         \and
        Edgard P. M. Amorim 
}


\institute{A. C. Orthey Jr \at
              Departamento de F\'isica, Universidade do Estado de Santa Catarina, 89219-710,
Joinville, SC, Brazil \\
           \and
           E. P. M. Amorim \at
              Departamento de F\'isica, Universidade do Estado de Santa Catarina, 89219-710,
Joinville, SC, Brazil\\
              \email{edgard.amorim@udesc.br}
}

\date{Received: date / Accepted: date}

\maketitle

\begin{abstract}
We study the entanglement between the internal (spin) and external (position) degrees of freedom of the one-dimensional discrete time quantum walk starting from local and delocalized initial states whose time evolution is driven by Hadamard and Fourier coins. We obtain the dependence of the asymptotic entanglement with the initial dispersion of the state and establish a way to connect the asymptotic entanglement between local and delocalized states. We find out that the delocalization of the state increases the number of initial spin states which achieves maximal entanglement from two states (local) to a continuous set of spin states (delocalized) given by a simple relation between the angles of the initial spin state. We also carry out numerical simulations of the average entanglement along the time to confront with our analytical results.


\end{abstract}

\section{Introduction}\label{intro}

Quantum random walks \cite{aharonov1993quantum,kempe2003quantum} or
quantum walks (QW) are the quantum counterparts of the classical
random walks. The quantum walker is a qubit, a particle with an
internal degree of freedom (spin) on a regular lattice where each
site represents an external degree of freedom (position). The
dynamic evolution of a QW starting from an initial state is given by
successive applications of a unitary time evolution operator, which
is constituted by a quantum coin and a conditional displacement.
While the quantum coin operates only on the qubit's internal degree
of freedom leaving it in a superposition of spin states, the
conditional displacement operator associates internal and external
degrees of freedom displacing the qubit from a site to the neighbor
right (left) site if its spin state is up (down).

The particular dynamics of a QW dramatically changes the spreading
behavior of the quantum state leading to a quadratic gain in its
dispersion (ballistic behavior), and it creates entanglement between
spin and position states. For these features, QW have been widely
investigated opening new branches of research in physics,
computational science and engineering \cite{venegas2012quantum}.
They offer insights for building quantum search algorithms
\cite{shenvi2003quantum}, for understanding the efficiency of energy
transfer in the photosynthesis \cite{engel2007evidence}, for
performing universal computation in their continuous
\cite{childs2009universal} and discrete time
\cite{lovett2010universal} versions, for simulating Dirac-like
Hamiltonians \cite{chandrashekar2013two}, and they can be
implemented in several experimental platforms
\cite{wang2013physical}.

The long-time entanglement in QW has a strong dependence on their
initial conditions. The initial state of a QW could be an arbitrary
qubit placed in one position (delta-like or local) or distributed
over many positions (delocalized state). For a QW starting from a
local state, the maximal entanglement is reached asymptotically and
only for few specific initial spin states
\cite{abal2006quantum,abal2008erratum,salimi2012asymptotic,eryiugit2014time}
and also for two walkers \cite{alles2012maximal}. The maximal
entanglement is also achieved regardless of the initial state
through the introduction of a dynamic disorder along the QW, such as
a random quantum coin in each time step
\cite{vieira2013dynamically,vieira2014entangling}. There are few
papers about delocalized initial conditions in QW showing a rich
variety of spreading behavior highly dependent of the quantum coin
\cite{valcarcel2010tailoring,zhang2016creating} and their
entanglement content
\cite{romanelli2010distribution,romanelli2012thermodynamic,romanelli2014entanglement}.
However, none of these earlier works address the interplay between
delocalization and asymptotic entanglement for all initial spin
conditions and two kinds of coins (Hadamard and Fourier) or yet,
they do not show a comparison between the average entanglement
behavior along the time with their analytical results.

Our main focus here is the impact of the delocalization of the
initial state in QW regarding their asymptotic entanglement. We
perform all calculations also to the local state in order to
contrast with the delocalized cases. In this way this article is
organized as follows. In Sect.~\ref{sec:2}, we review the QW in
position and momentum space using a Hadamard coin, the
quantification of the entanglement in both spaces and we obtain a
general expression for the asymptotic entanglement. In
Sect.~\ref{sec:3}, we investigate the asymptotic entanglement for QW
starting from local and delocalized states (Gaussian and rectangular
states), we show a way to connect both kinds of asymptotic
entanglements and we confront our analytical results with numerical
calculations. In Sect.~\ref{sec:4}, a general conclusion is
pictured. Finally, in "Appendix" we extend the calculations of
Sect.~\ref{sec:2} and the results of Sect.~\ref{sec:3} for a QW
which evolves by means of a Fourier coin.

\section{Mathematical formalism}\label{sec:2}

The quantum walker state $\ket{\Psi}$ belongs to a Hilbert space $\mathcal{H}=\mathcal{H}_C\otimes\mathcal{H}_P$ where
$\mathcal{H}_C$ is a single qubit coin space, a two-dimensional complex vector space spanned by the spin states $\{\ket{\up}, \ket{\down}\}$ and $\mathcal{H}_P$ is the position space, an infinite-dimensional enumerable vector space spanned by a set of orthonormal
vectors ${\ket{j}}$ where the integer $j$ is the discrete position
of the qubit on a one-dimensional lattice. Thus, a general initial state is
\begin{equation}
\ket{\Psi(0)}=\sum_{j=-\infty}^{+\infty}\left[a(j,0)\ket{\up}+b(j,0)\ket{\down}\right]\otimes\ket{j},
\label{Psi_0}
\end{equation}
with the normalization condition,
\begin{equation}
\sum_{j=-\infty}^{+\infty}\left(|a(j,0)|^2+|b(j,0)|^2\right)=1.
\end{equation}

The time evolution of a QW state is written as $\ket{\Psi(t)}=U^t\ket{\Psi(0)}$ and the time evolution operator is
\begin{equation}
U=S.(C\otimes\mathbb{1}_P),
\end{equation}
where $\mathbb{1}_P$ is the identity in $\mathcal{H}_P$ and $C$ is
the quantum coin. The quantum coin acts over the spin states and
generates a superposition of them. Here, we employ the
Hadamard\footnote{In "Appendix" we extend the following calculations
for a Fourier coin.}, widely used as a quantum coin,
\begin{equation}
H=\dfrac{1}{\sqrt{2}}\begin{pmatrix}
1 & 1 \\
1 & -1
\end{pmatrix}.
\label{Hadamard_op}
\end{equation}
The conditional displacement operator
\begin{equation}
S=\sum_j(\ket{\up}\bra{\up}\otimes\ket{j+1}\bra{j}+\ket{\down}\bra{\down}\otimes\ket{j-1}\bra{j}),
\end{equation}
moves the qubit to the right or left conditioned to its internal state, i.e., from site $j$ to $j+1$ ($j-1$) if its spin is up (down), which creates entanglement between spin and position along the time evolution.

The total initial state $\ket{\Psi(0)}$ is pure and since the time evolution is unitary, $\ket{\Psi(t)}$ remains pure. This fact allows us to quantify the entanglement between spin and position by means of von Neumann entropy
\begin{equation}
S_E(\rho(t))=-Tr(\rho_C(t)\log_2\rho_C(t)),
\end{equation}
of the partially reduced coin (spin) state $\rho_C(t)=Tr_P(\rho(t))$ \cite{bennett1996concentrating}, where $\rho(t)=\ket{\Psi(t)}\bra{\Psi(t)}$ and $Tr_P(\cdot)$ is the trace over the positions. Therefore, we can write %
\begin{equation}
\rho_C(t)=A(t)\ket{\up}\bra{\up}+B(t)\ket{\up}\bra{\down}+B(t)^*\ket{\down}\bra{\up}+C(t)\ket{\down}\bra{\down},
\end{equation}
with $A(t)=\sum_{j}|a(j,t)|^2$, $B(t)=\sum_{j}a(j,t)b^*(j,t)$ and $C(t)=1-A(t)$ due to the normalization condition and $B(t)^*$ is the complex conjugate of $B(t)$. After diagonalizing $\rho_C(t)$, we obtain
\begin{equation}
S_E(\rho(t))=-\lambda_+(t)\log_2\lambda_{+}(t)-\lambda_{-}(t)\log_2\lambda_{-}(t),
\label{SE_rho}
\end{equation}
where the eigenvalues of $\rho_C(t)$ are
\begin{equation}
\lambda_{\pm}(t)=\dfrac{1}{2}{\pm}\sqrt{\dfrac{1}{4}-A(t)(1-A(t))+|B(t)|^2}.
\label{lambda}
\end{equation}
$S_E(\rho(t))$ ranges from $0$ for separable states up to $1$ for maximal entanglement condition. In the context of a numerical simulation, the expression \eqref{SE_rho} with \eqref{lambda} are used to calculate the entanglement along the time.

To quantify the entanglement in the asymptotic limit for $t\rightarrow\infty$, we need to consider the dual k-space $\mathcal{\tilde{H}}_k$ spanned by the Fourier transformed vectors $\ket{k}=\sum_{j}e^{ikj}\ket{j}$ with $k\in[-\pi,\pi]$, where the initial state \eqref{Psi_0} is,
\begin{equation}
\ket{\tilde{\Psi}(0)}=\int_{-\pi}^{\pi}\dfrac{dk}{2\pi}[\tilde{a}_k(0)\ket{\up}+\tilde{b}_k(0)\ket{\down}]\otimes\ket{k},
\label{Psitil_0}
\end{equation}
and the corresponding initial amplitudes are
\begin{equation}
\begin{array}{rl}
\tilde{a}_k(0)&=(\bra{\up}\otimes\bra{k})\ket{\tilde{\Psi}(0)}=\sum_j e^{-ikj}a(j,0),\\
\tilde{b}_k(0)&=(\bra{\down}\otimes\bra{k})\ket{\tilde{\Psi}(0)}=\sum_j e^{-ikj}b(j,0).
\label{abk0}
\end{array}
\end{equation}
The time evolution operator $U$ can be rewritten as \cite{nayak2001one},
\begin{equation}
U_H=\dfrac{1}{\sqrt{2}}\begin{pmatrix}
e^{-ik} & e^{-ik} \\
e^{ik} & -e^{ik}
\end{pmatrix}, \label{evolutionk_op}
\end{equation}
once the conditional displacement operator $S$ is diagonal in the Fourier representation,
\begin{equation}
S_k=\ket{\up}\bra{\up}\otimes e^{-ik}\ket{k}\bra{k}+ \ket{\down}\bra{\down}\otimes e^{ik}\ket{k}\bra{k}.
\end{equation}
Let us introduce the state $\ket{\Phi_k(t)}=\braket{k|\tilde{\Psi}(t)}$, such that the one-step time evolution is $\ket{\Phi_k(t+1)}=U_H\ket{\Phi_k(t)}$. By calculating the eigenvectors of $U_H$, we obtain
\begin{equation}
\ket{\Phi_k^{\pm}}=\dfrac{\left[1+\cos^2k\mp(\cos k \sqrt{1+\cos^2k}) \right]^{-1/2}}{\sqrt{2}}
\begin{pmatrix}
e^{-ik} \\
\pm\sqrt{2}e^{-i\omega_k}-e^{-ik}
\end{pmatrix},
\label{eigenvectors}
\end{equation}
and the eigenvalues are $\pm e^{\mp i\omega_k}$ and the frequency $\omega_k$ given by $\sin\omega_k=\sin k/\sqrt{2}$ with $\omega_k\in[-\pi/2,\pi/2]$ \cite{nayak2001one,abal2006quantum}.

The time evolution starting from an initial state $\ket{\Phi_k(0)}$ is $\ket{\Phi_k(t)}=(U_H)^t\ket{\Phi_k(0)}$ and through the spectral decomposition of $U_H$, it could be written in the following way,
\begin{equation}
\ket{\Phi_k(t)}=e^{-i\omega_k t}\braket{\Phi_k^{+}|\Phi_k(0)}\ket{\Phi_k^{+}}+(-1)^te^{i\omega_kt}\braket{\Phi_k^{-}|\Phi_k(0)}\ket{\Phi_k^{-}}.
\label{Phikt}
\end{equation}
This expression allows us to get the time evolution of each spin amplitude from the corresponding initial amplitude. Therefore, calculating the integrals
\begin{align}
A(t) &= \displaystyle\int_{-\pi}^{\pi}\dfrac{dk}{2\pi}|\tilde{a}_k(t)|^2,
\label{A}\\
B(t) &= \displaystyle\int_{-\pi}^{\pi}\dfrac{dk}{2\pi}\tilde{a}_k(t)\tilde{b}_k^*(t),
\label{B}
\end{align}
and inserting them in \eqref{SE_rho} using \eqref{lambda}, we have the entanglement as function of the amplitudes in the momentum space.

To obtain the asymptotic entanglement, we must take the limit for $t\rightarrow\infty$ in \eqref{Phikt}, then the time dependence vanishes in \eqref{A} and \eqref{B}, thereby we have
\begin{align}
\overline{A}=&\int_{-\pi}^{\pi}\dfrac{dk}{2\pi} \left\{\dfrac{1}{2 (3+\cos (2 k))} \left[ 4|\tilde{a}_k(0)|^2+\tilde{a}_k^*(0)\tilde{b}_k(0)+\tilde{a}_k(0)\tilde{b}_k^*(0)+2|\tilde{b}_k(0)|^2  \right.\right.\nonumber\\
& +\left.\left. \left(\tilde{a}_k^*(0)\tilde{b}_k(0)+\tilde{a}_k(0)(2\tilde{a}_k^*(0)+\tilde{b}_k^*(0))\right)\cos(2k)\right.\right.\nonumber\\
&\left.\left.-i\left(\tilde{a}_k^*(0)\tilde{b}_k(0)-\tilde{a}_k(0)\tilde{b}_k^*(0)\right)\sin(2k) \right]\right\},
\label{A_asym}\\
\overline{B}&=\int_{-\pi}^{\pi}\dfrac{dk}{2\pi}\left\{\dfrac{\cos k-i\sin k}{3+\cos(2k)}\left[ \left( \tilde{a}_k^*(0)\tilde{b}_k(0)+\tilde{a}_k(0)\tilde{b}_k^*(0)\right.\right.\right.\nonumber\\
&\left.\left.\left.+|\tilde{a}_k(0)|^2-|\tilde{b}_k(0)|^2 \right)\cos k\right.\right.-\left.\left.
i\left(\tilde{a}_k^*(0)\tilde{b}_k(0)-\tilde{a}_k(0)\tilde{b}_k^*(0)\right)\sin k \right] \right\},
\label{B_asym}
\end{align}
where $\overline{A}$ is $A(t\rightarrow\infty)$. After inserting
these Eqs. \eqref{A_asym} and \eqref{B_asym} into \eqref{SE_rho}, we
have \cite{ide2001entanglement}
\begin{align}
\overline{S}_E(\overline{\Delta})&=-(\overline{\lambda}_+)\log_2(\overline{\lambda}_+)-(\overline{\lambda}_-)\log_2(\overline{\lambda}_-),
\label{SE_asym}
\end{align}
with $\overline{\lambda}_{\pm}=(1\pm\sqrt{\overline{\Delta}})/2$, a general expression to the asymptotic entanglement in terms of a characteristic function,
\begin{equation}
\overline{\Delta}(\overline{A},\overline{B})=1-4\left[\overline{A}(1-\overline{A})+|\overline{B}|^2\right],
\label{Delta}
\end{equation}
which contains all the information about the initial state, since $\overline{A}$ and $\overline{B}$ are functions of the initial amplitudes in the k-space.

\section{Results} \label{sec:3}

\begin{figure*}
\centering
\includegraphics[width=84mm]{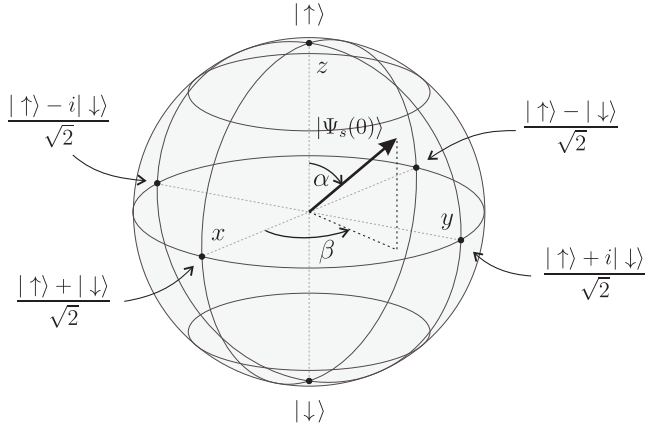}
\caption{Bloch sphere representation of all possible initial spin states from eq. \eqref{Psi_s}. Here, $\alpha$ and $\beta$ are the polar and azimuthal angles, respectively, from standard spherical coordinates. Some spin states are detached.}
\label{fig:1}
\end{figure*}

Our calculations starting from a local state followed by two delocalized states. In all cases, we consider a general spin state,
\begin{equation}
\ket{\Psi_s(0)}=\cos\frac{\alpha}{2}\ket{\up}+e^{i\beta}\sin\frac{\alpha}{2}\ket{\down},
\label{Psi_s}
\end{equation}
represented in the Bloch sphere \cite{nielsen2010quantum} in Fig.~\ref{fig:1}.

\subsection{Local state}

Consider the spin state \eqref{Psi_s} and the probability distribution $|\Psi_L(0)|^2=\delta(j)$ in \eqref{Psi_0} to get a local state,
\begin{equation}
\ket{\Psi_L(0)}=\ket{\Psi_s(0)}\otimes\ket{0}.
\label{Psi_0_Local}
\end{equation}
Theses amplitudes can be rewritten in the k-space by using
\eqref{abk0} as $\tilde{a}_k(0)=\cos(\alpha/2)$ and
$\tilde{b}_k(0)=e^{i\beta}\sin(\alpha/2)$ and inserting them in
\eqref{A_asym} and \eqref{B_asym}, after integrating both equations
we reach
\begin{align}
\label{A_asym_local}\overline{A} &=\dfrac{1}{2}+\left(\dfrac{1}{4}-f\right)\left(\cos\alpha+\sin\alpha\cos\beta \right),\\
\label{B_asym_local}\overline{B} &=\overline{A}-\dfrac{1}{2}+ 2if\sin\alpha\sin\beta,
\end{align}
where $f=-(1-\sqrt{2})/4$. Thus, by inserting them into \eqref{Delta}, we have a characteristic function for local state,
\begin{equation}
\overline{\Delta}_H(\alpha,\beta)=(3-2\sqrt{2})[1+\sin(2\alpha)\cos\beta].
\label{Delta_Local}
\end{equation}

\begin{figure*}
\centering
\includegraphics[width=84mm]{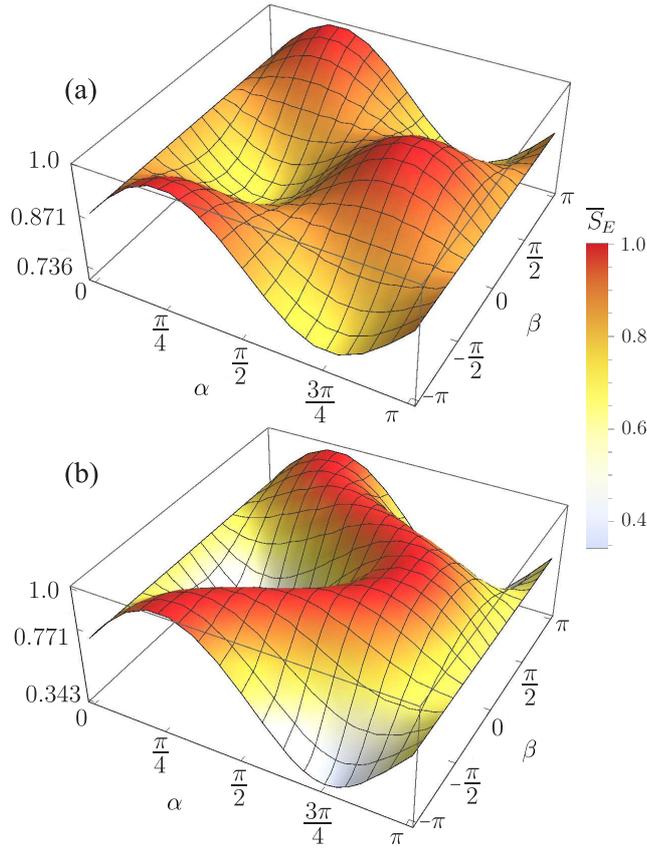}
\caption{Asymptotic entanglement $\overline{S}_E$ starting from (a) local state \eqref{Psi_0_Local} and (b) Gaussian state \eqref{Psi_0_Gauss} with $\sigma_0=1$ as function of the angles $\alpha$ and $\beta$ from the initial spin states amplitudes.}
\label{fig:2}
\end{figure*}

Figure~\ref{fig:2} (a) shows the asymptotic entanglement calculated from \eqref{Delta_Local} in \eqref{SE_asym} starting from a local state for any initial spin state. The maximum entanglement is $\overline{S}_E=1.0$ for $(\alpha,\beta)=(3\pi/4,0)$ and $(\pi/4,\pi)$. The minimum entanglement is $\overline{S}_E\sim 0.736$ for $(\alpha,\beta)=(\pi/4,0)$ and $(3\pi/4,\pi)$ \cite{abal2006quantum,abal2008erratum}.

\subsection{Delocalized states}

Let us consider a Gaussian probability distribution,
\begin{equation}
|\Psi_G(0)|^2=\dfrac{\text{exp}(-j^2/2\sigma_0^2)}{\sqrt{2\pi\sigma_0^2}},
\end{equation}
with an initial dispersion $\sigma_0$ and since $\ket{\Psi_s(0)}$ is the initial spin state, the discrete Gaussian state\footnote{The Gaussian states were defined between $[-1000,1000]$, i.e., for a $\sigma_0=30$ the discrete sum of the normalization condition gives an error below of $0.001\%$.} could be written as,
\begin{equation}
\ket{\Psi_G(0)}=\sum_{j=-\infty}^{+\infty}\ket{\Psi_s(0)}\otimes\frac{\text{exp}\left(-j^2/4\sigma_0^2\right)}{(2\pi\sigma_0^2)^{\frac{1}{4}}}\ket{j}.
\label{Psi_0_Gauss}
\end{equation}
In the same way as in the previous case, using \eqref{abk0} to rewrite the Gaussian amplitudes in the k-space and changing the sum of amplitudes in $j$ by their integration\footnote{Since the numerical difference between the discrete sum and integration of the amplitudes with $\sigma_0=1$ is around $10^{-4}$, and this difference is even smaller for larger $\sigma_0$.} in $x$, we have
\begin{equation}
\begin{pmatrix}
\tilde{a}_k(0) \\
\tilde{b}_k(0)
\end{pmatrix}= \int\limits_{-\infty}^{+\infty}\frac{\text{exp}\left( -x^2/(4\sigma_0^2)-ikx \right)}{\left( 2\pi\sigma_0^2 \right)^{\frac{1}{4}}}dx \ket{\Psi_s(0)}.
\label{ab_k_int_Gauss}
\end{equation}
After integrating \eqref{ab_k_int_Gauss} \cite{abramowitz1964handbook}, the imaginary terms vanish, therefore, we have
\begin{equation}
(\tilde{a}_k(0),\tilde{b}_k(0))^T= \left(8\pi\sigma_0^2\right)^{\frac{1}{4}}e^{-k^2\sigma_0^2}\ket{\Psi_s(0)},
\label{ab_k_Gauss}
\end{equation}
are the Gaussian initial amplitudes in the k-space. Inserting
\eqref{ab_k_Gauss} in \eqref{A_asym} and \eqref{B_asym}, after
numerical integration we obtain the same Eqs. \eqref{A_asym_local} e
\eqref{B_asym_local}, however $f$ is a function of initial
dispersion $\sigma_0$ given by $f(\sigma_0)=\epsilon/\sigma_0^2$,
where the constant $\epsilon\sim0.0327$. Then, the characteristic
function for Gaussian states is
\begin{align}
\overline{\Delta}_H(\alpha, \beta, \sigma_0)&=
\dfrac{1}{2}\left(1-4f(\sigma_0)\right)^2\left[\cos\alpha+\sin\alpha\cos\beta\right]^2\nonumber\\
& +\left(4f(\sigma_0)\right)^2[\sin\alpha\sin\beta]^2.
\label{Delta_Gauss}
\end{align}

Figure~\ref{fig:2} (b) shows the asymptotic entanglement starting from a Gaussian state obtained from \eqref{Delta_Gauss} with $\sigma_0=1$ in \eqref{SE_asym} for any initial spin state. There is a continuous range of spin states with maximum entanglement, and the minimum entanglement is $\overline{S}_E\sim0.343$, for the same angles of local case.

\begin{figure*}
\centering
\includegraphics[width=\linewidth]{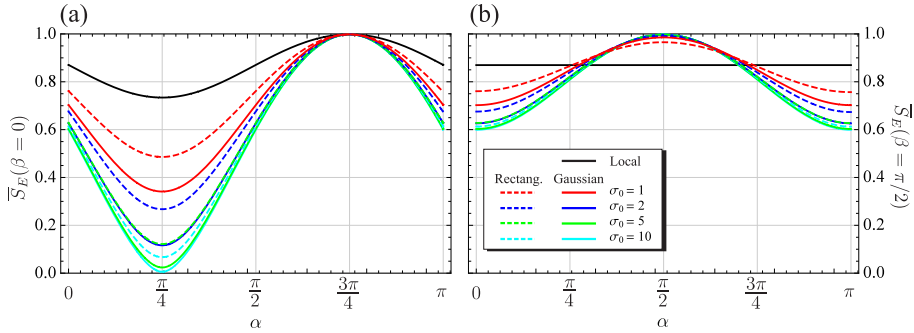}
\caption{Asymptotic entanglement $\overline{S}_E$ for (a) $\beta=0$ and (b) $\beta=\pi/2$ for $\alpha\in[0,\pi]$.}
\label{fig:3}
\end{figure*}

To investigate if the asymptotic entanglement has any dependence on the shape of the initial delocalized state, we also study a QW starting from a rectangular state. The rectangular probability distribution is $|\Psi_R(0)|^2=1/(2a+1)$ if $-a\leq j\leq a$ and $0$, otherwise, then a rectangular state is given by,
\begin{equation}
\ket{\Psi_R(0)}=\sum_{j=-a}^{a}\ket{\Psi_s(0)}\otimes\dfrac{1}{\sqrt{2a+1}}\ket{j}.
\end{equation}
We obtain their amplitudes in the k-space using \eqref{abk0} such as the previous cases. After the sum we have,
\begin{equation}
\begin{pmatrix}
\tilde{a}_k(0) \\
\tilde{b}_k(0)
\end{pmatrix} = \left[\frac{\sin(ka+k)}{\tan(k/2)}-\cos(ka+k)\right]\frac{\ket{\Psi_s(0)}}{\sqrt{2a+1}},
\label{ab_k_int_Rectang}
\end{equation}
and by inserting these amplitudes in \eqref{A_asym} and
\eqref{B_asym} and performing a numerical integration, we have the
same Eqs. \eqref{A_asym_local} and \eqref{B_asym_local}, which leads
us to the same characteristic function obtained from the Gaussian
state in \eqref{Delta_Gauss}, however in this case $f(a)=\epsilon/a$
where $\epsilon\sim0.0684$ with $a=(\sqrt{12\sigma_0^2+1}-1)/2$.

The characteristic function has the same analytical form \eqref{Delta_Gauss} for both delocalized states. For large initial dispersion $\sigma_0\gg1$, we reach
\begin{equation}
\overline{\Delta}_H(\alpha,\beta)\sim (1/2)[\cos\alpha+\sin\alpha\cos\beta]^2.
\end{equation}
Since $\overline{S}_E=1$ for $\overline{\Delta}=0$, we have
\begin{equation}
\cos\beta=-\cot\alpha,
\label{max_entang_cond}
\end{equation}
a maximal entanglement condition for high delocalization. On the
opposite way, for $\alpha=\pi/4$ and $\beta=0$,
$\overline{S}_E\rightarrow 0$\footnote{The minimum entanglement
condition for Gaussian and rectangular states obeys a power law
$\overline{S}_E\sim 0.3463\sigma_0^{-1.59}$ and $\overline{S}_E\sim
0.4874\sigma_0^{-0.853}$, respectively, both obtained by curve
fitting.} for larger $\sigma_0$, as we can see in Fig.~\ref{fig:3}
which compares the asymptotic entanglement for local, Gaussian and
rectangular states for (a) $\beta=0$ and (b) $\beta=\pi/2$. The
delocalized states have maximum entanglement for both values of
$\beta$ in agreement with \eqref{max_entang_cond}.

\begin{figure*}
\centering
\includegraphics[width=\linewidth]{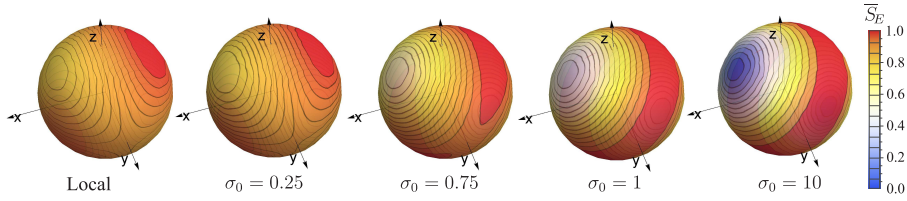}
\caption{Asymptotic entanglement $\overline{S}_E$ over the Bloch sphere starting from local to Gaussian states with $\sigma_0=0.25$, $0.75$, $1$ and $10$ (left to right). The azimuthal angle $\beta$ is around $\hat{z}$ and $\alpha$ is the polar angle. Red (blue) regions show high (low) entanglement rates.}
\label{fig:4}
\end{figure*}

\paragraph{Connecting local and delocalized states} We have obtained an analytical expression for the asymptotic entanglement for local and delocalized states using the same theoretical framework, although there is a lack of connection between these two cases. In attempt to fulfill this requirement, we have to impose a renormalization of the spin amplitudes $\tilde{a}_k(0)$ and $\tilde{b}_k(0)$ for a Gaussian state, i.e., $\ket{\tilde{\Psi}_G(0)}/\text{Erf}(\sqrt{2}\pi\sigma_0)^{1/2}$ where $\text{Erf}(z)=(2/\sqrt{\pi})\int_{0}^{z}e^{-u^2}du$ in order to keep it normalized for $0<\sigma_0<1$. After an extensive numerical integration, we get approximate expressions $\overline{A}$ and $\overline{B}$ analogous to the previous cases, although $f$ is
\begin{equation}
f(\sigma_0)=0.0365\left\{\pi/2-\text{ArcTan}[3.937(\sigma_0-0.8)]\right\},
\end{equation}
such that, $f(\sigma_0\rightarrow 0)\sim -(1-\sqrt{2})/4$ to recover the local case. Figure~\ref{fig:4} shows $\overline{S}_E$ over the Bloch sphere, starting from local to Gaussian states enhancing the high entanglement region from two spots up to a strip around the Bloch sphere with a maximum entanglement which obeys \eqref{max_entang_cond}.

\subsection{Numerical simulations}

On the one hand, the long-time entanglement in the position space can be obtained through an iterative computational way after many time steps calculated by \eqref{SE_rho}. The asymptotic entanglement, on the other hand, can be straightforward calculated by Fourier analysis as showed above. However, what is the general behavior of the entanglement along the time evolution? How many steps are necessary to be close enough to the asymptotic entanglement? In order to enlighten these questions, first we perform a numerical simulation of the average entanglement $\braket{S_E(t)}$ along the time for all cases. Second, for delocalized cases, we make a comparison between the average asymptotic entanglement and the average entanglement numerically evaluated as function of the initial dispersion.

\begin{figure*}
\centering
\includegraphics[width=84mm]{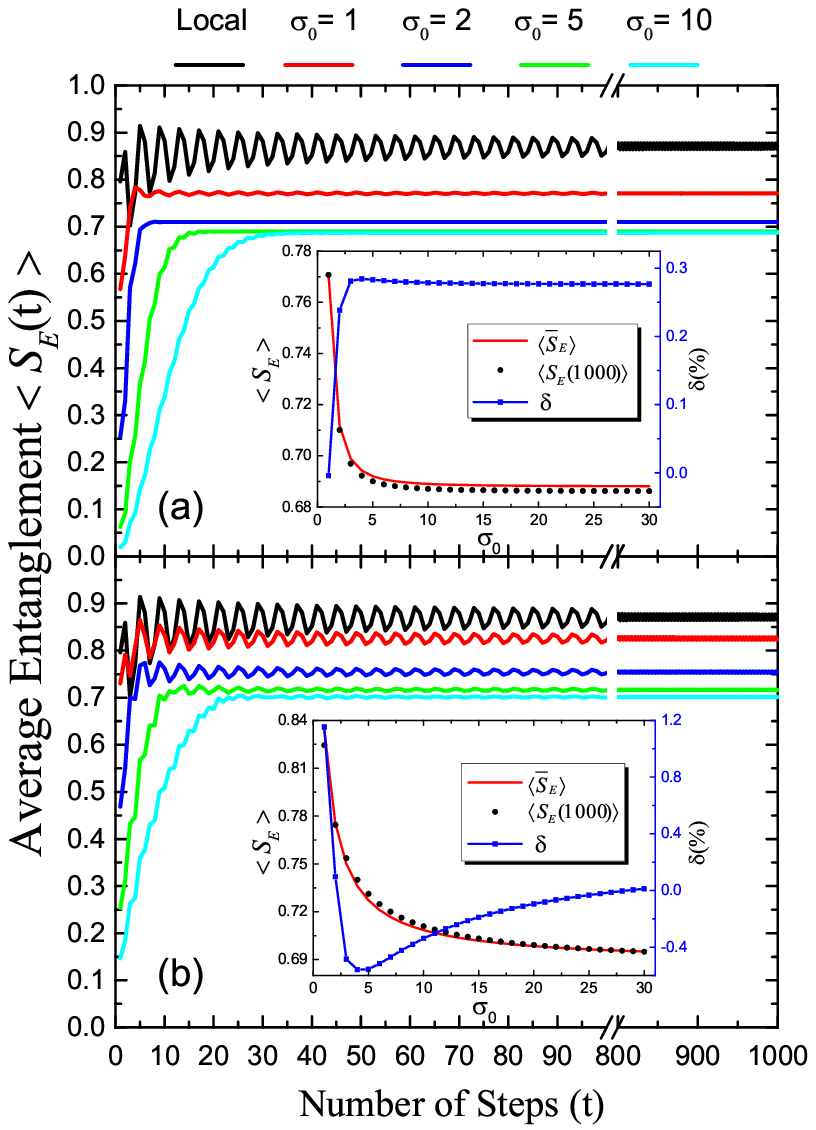}
\caption{The average entanglement $\braket{S_E(t)}$ was obtained over 2016 initial spin states from $(\alpha,\beta)=(0,0)$ up to $(\pi, 2\pi)$ in independent increments of $0.1$. The black curve shows the local case. The colored curves are (a) Gaussian and (b) rectangular for $\sigma_0=1$, $2$, $5$, and $10$. Inset: Average entanglement after 1000 time steps $\braket{S_E(1000)}$ (black dots) and average asymptotic entanglement $\braket{\overline{S}_E}$ (red curve) obtained from Fourier analysis as function of initial dispersion $\sigma_0$. In blue, the percentage difference $\delta(\%)$ between both calculations.}
\label{fig:5}
\end{figure*}

Figure~\ref{fig:5} shows the average entanglement $\braket{S_E(t)}$
as function of the time steps starting from a (a) Gaussian and (b)
rectangular states, both for distinct values of initial dispersion
$\sigma_0$ together with the local state. In both insets, we compare
the asymptotic entanglement $\braket{\overline{S}_E}$ with the
$\braket{S_E(1000)}$ numerically obtained after $1000$ time steps as
function of the initial dispersion $\sigma_0$ being the difference
smaller than $0.3\%$ and $1.2\%$, respectively, for (a) Gaussian and
(b) rectangular cases. The average asymptotic entanglement
$\braket{\overline{S}_E}$ has a dependence on the initial dispersion
$\sigma_0$ being always greater than $\braket{\overline{S}_E}\sim
0.688$ for both delocalized cases. Fitting the asymptotic curve
(red) gives us a $\sigma_0^2$ decay,
$\braket{\overline{S}_E}_G(\sigma_0)\sim0.0835\sigma_0^{-2}+0.688$
for Gaussian states and a $\sigma_0$ decay,
$\braket{\overline{S}_E}_R(\sigma_0)\sim0.1205\sigma_0^{-1}+0.688$
for rectangular cases, both obtained by means of numerical
integration.

\section{Conclusion} \label{sec:4}

The main aspect emphasized here is the effect of delocalization of
the initial state regarding the generation of entanglement. We have
obtained the asymptotic entanglement in QW starting from local and
two kind of delocalized states, as well as a way to connect them
through analytical and numerical analysis. For delocalized states,
we found out a simple relation between the initial angles of the
spin amplitudes \eqref{max_entang_cond}, which always leads to the
maximal entanglement for Hadamard and Fourier (see "Appendix")
walks. This relation for both cases expands our knowledge on maximal
entanglement achievement from two specific spin states (local) up to
a continuous set of initial spin states (delocalized). It is worth
mentioning that the QW from delocalized states considered here have
ballistic spreading since their evolution is ordered, as opposed to
the diffusive behavior for disordered scenarios
\cite{vieira2013dynamically,vieira2014entangling}.

Finally, we hope that our findings help improve the development of entanglement generation protocols and the experimentalists can test our results in different experimental platforms \cite{wang2013physical}.

\begin{acknowledgements}
ACO thanks CAPES (Brazilian Agency for the Improvement of Personnel of Higher Education) for the grant and EPMA thanks Janice Longo for her careful reading of the manuscript.
\end{acknowledgements}

\appendix

\section{Fourier walk} \label{sec:Fourier}

The Fourier (or Kempe) is a balanced coin such as the Hadamard,
\begin{equation}
F=\dfrac{1}{\sqrt{2}}\begin{pmatrix}
1 & i\\
i & 1
\end{pmatrix},
\end{equation}
however it generates a symmetric probability distribution without phase difference between initial spin states ($\beta=0$) unlike the Hadamard ($\beta=\pi/2$) \cite{kempe2003quantum}. The time evolution operator in the k-space considering a Fourier coin is
\begin{equation}
U_F=\dfrac{1}{\sqrt{2}}\begin{pmatrix}
e^{-ik} & ie^{-ik} \\
ie^{ik} & e^{ik}
\end{pmatrix},
\end{equation}
with eigenvectors $\ket{\Phi_k^{\pm}}$ given by
\begin{equation}
\ket{\Phi_k^{\pm}}=\dfrac{1}{\sqrt{2}}\left( 1+\sin^2k\pm\sin k\sqrt{1+\sin^2k}\right)^{-\frac{1}{2}}\begin{pmatrix}
-e^{-ik}\left(\sin k\pm \sqrt{1+\sin^2k} \right)\\
1
\end{pmatrix},
\label{eigenvectorsF}
\end{equation}
and the eigenvalues are
\begin{equation}
\lambda^{\pm}=\dfrac{1}{\sqrt{2}}\left(\cos k\mp i\sqrt{1+\sin^2 k} \right).
\end{equation}
If we define a frequency $\omega_k$ such that $\cos(\omega_k)=\cos k/\sqrt{2}$ implies that
\begin{equation}
\lambda^{\pm}=\cos(\omega_k)\mp i\sin(\omega_k)=e^{\mp i\omega_k}.
\end{equation}
At this point, we are able to write the time evolution $\ket{\Phi_k(t)}=(U_F)^t\ket{\Phi_k(0)}$ by means of a spectral decomposition of $U_F$
\begin{align}
\ket{\Phi_k(t)}&=e^{-i\omega_k t}\braket{\Phi^+_k|\Phi_k(0)}\ket{\Phi^+_k}+e^{i\omega_k t} \braket{\Phi^-_k|\Phi_k(0)}\ket{\Phi^-_k}.
\label{spec_decomp}
\end{align}
In order to make it easier to manipulate, let us write
\begin{align}
A^{\pm}_k &= \dfrac{1}{2}\left( 1+\sin^2k\pm\sin k\sqrt{1+\sin^2k}\right)^{-1},\\
u^{\pm}_k &= -e^{-ik}\left(\sin k\pm \sqrt{1+\sin^2k}\right),
\end{align}
in the eigenvectors \eqref{eigenvectorsF}, and since the initial state is $\ket{\Phi_k(0)}=\left(\tilde{a}_k(0),\tilde{b}_k(0)\right)^T$ therefore we have
\begin{align}
\ket{\Phi_k(t)}&=
e^{-i\omega_k t}A^+_k\left((u^+_k)^*\tilde{a}_k(0)+\tilde{b}_k(0)\right)\begin{pmatrix}
u^+_k\\
1
\end{pmatrix} \nonumber\\
&+e^{i\omega_k t}A^-_k\left((u_k^-)^*\tilde{a}_k(0)+\tilde{b}_k(0)\right)
\begin{pmatrix}
u^-_k\\
1
\end{pmatrix} ,
\end{align}
Let us consider $\ket{\Phi_k(t)}=\left(\tilde{a}_k(t),\tilde{b}_k(t)\right)^T$ where
\begin{align}
a_k(t)&= e^{-i\omega_k t}A^+_k\left(|u^+_k|^2\tilde{a}_k(0)+u^+_k \tilde{b}_k(0)\right)+e^{i\omega_k t}A^-_k\left(|u_k^-|^2\tilde{a}_k(0)+u^-_k \tilde{b}_k(0)\right), \\
b_k(t)&= e^{-i\omega_k t}A^+_k\left((u^+_k)^*\tilde{a}_k(0)+\tilde{b}_k(0)\right) +e^{i\omega_k t}A^-_k\left((u_k^-)^*\tilde{a}_k(0)+\tilde{b}_k(0)\right).
\end{align}
Since $\tilde{b}_k(t)$ is the simplest, we can use it to calculate the entanglement through the expressions,
\begin{align}
C(t)&=\int_{-\pi}^{\pi}\dfrac{dk}{2\pi}|\tilde{b}_k(t)|^2,\\
B(t)&=\int_{-\pi}^{\pi}\dfrac{dk}{2\pi}\tilde{a}_k(t)\tilde{b}^*_k(t).
\end{align}
When $t\rightarrow+\infty$, the time-dependent terms vanish \cite{nayak2001one}, leading to
\begin{align}
\overline{C}&= \int_{-\pi}^{\pi}\dfrac{dk}{2\pi} \left\{ (A^+_k)^2\left[ |u_k^+|^2|\tilde{a}_k(0)|^2+(u^+_k)^*\tilde{a}_k(0)\tilde{b}^*_k(0)+ u^+_k\tilde{a}_k(0)\tilde{b}_k(0)+ |\tilde{b}_k(0)|^2\right]\right. \nonumber\\
& +\left.(A^-_k)^2\left[ |u_k^-|^2|\tilde{a}_k(0)|^2+(u^-_k)^*\tilde{a}_k(0)\tilde{b}^*_k(0)+ u^-_k\tilde{a}_k(0)\tilde{b}_k(0)+ |\tilde{b}_k(0)|^2\right]\right\},
\label{C_asym_fourier}\\
\overline{B}&= \int_{-\pi}^{\pi}\dfrac{dk}{2\pi} \left\{ (A^+_k)^2\left[ |u_k^+|^2\left(u^+_k|\tilde{a}_k(0)|^2+ \tilde{a}_k(0)\tilde{b}^*_k(0)\right)+(u^+_k)^2\tilde{a}_k(0)\tilde{b}_k(0)+ u_k^+|\tilde{b}_k(0)|^2\right] \right.\nonumber\\
&\left. +(A^-_k)^2\left[ |u_k^-|^2\left(u^-_k|\tilde{a}_k(0)|^2+ \tilde{a}_k(0)\tilde{b}^*_k(0)\right)+ (u^-_k)^2\tilde{a}_k(0)\tilde{b}_k(0)+ u_k^-|\tilde{b}_k(0)|^2\right]\right\}.
\label{B_asym_fourier}
\end{align}
After inserting the Eqs. \eqref{C_asym_fourier} and
\eqref{B_asym_fourier} in the entanglement Eq. \eqref{SE_rho}, we
reach the asymptotic entanglement given by \eqref{SE_asym} as
function of a characteristic function \eqref{Delta}.

\paragraph{Local Initial State} Let us consider the initial local state given by k-space amplitudes
$\tilde{a}_k(0)=\cos(\alpha/2)$ and $\tilde{b}_k(0)=e^{i\beta}\sin(\alpha/2)$ and inserting them in \eqref{C_asym_fourier} and \eqref{B_asym_fourier} to obtain the characteristic function for the Fourier walk,
\begin{equation}
\overline{\Delta}_F(\alpha,\beta)=(3-2\sqrt{2})[1-\sin(2\alpha)\sin\beta],
\label{delta_f}
\end{equation}
which satisfies $\overline{\Delta}_F(\alpha,\beta-\pi/2)=\overline{\Delta}_H(\alpha,\beta)$ and the maximum entanglement is reached for $(\alpha,\beta)=(\pi/4,\pi/2)$ and $(3\pi/4,-\pi/2)$.

\begin{figure}
\centering
\includegraphics[width=\linewidth]{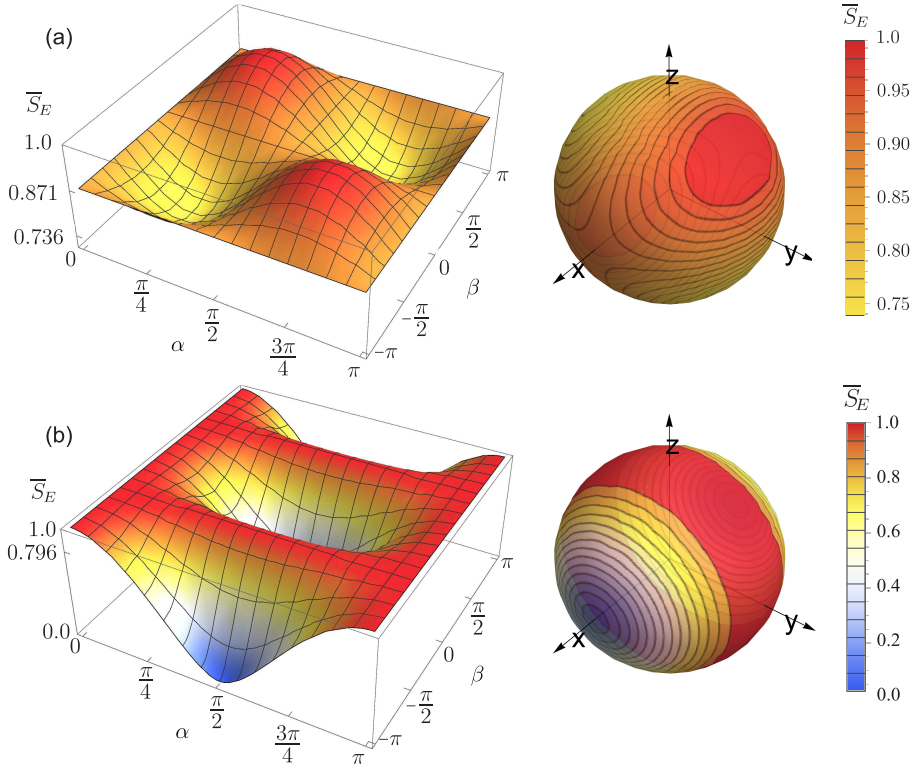}
\caption{Asymptotic entanglement $\overline{S}_E$ as function of the angles $\alpha$ and $\beta$ from the initial spin amplitudes and over the Bloch sphere of a Fourier walk from a (a) local state with $\braket{\overline{S}_E}=0.871$ and $0.736$, such as the Hadamard walk and (b) Gaussian state ($\sigma_0=10$) with $\braket{\overline{S}_E}=0.796$ and $\sim 0$, respectively for the average and minimum entanglement. Red (blue) regions show high (low) entanglement rates.}
\label{fig:6}
\end{figure}

\paragraph{Gaussian Initial State} Inserting the Gaussian k-space amplitudes \eqref{ab_k_Gauss} in \eqref{C_asym_fourier} and \eqref{B_asym_fourier} leads us to the characteristic function of Fourier walk starting from a Gaussian state,
\begin{align}
\overline{\Delta}_F(\alpha, \beta, \sigma_0)&=
\dfrac{1}{2}\left(1-4f(\sigma_0)\right)^2\left[\cos\alpha-\sin\alpha\sin\beta\right]^2
+\left(4f(\sigma_0)\right)^2[\sin\alpha\cos\beta]^2,
\label{Delta_Gauss_f}
\end{align}
that also satisfies $\overline{\Delta}_F(\alpha,\beta-\pi/2)=\overline{\Delta}_H(\alpha,\beta)$. However, for this case $f(\sigma_0)\rightarrow 1/4$ when $\sigma_0\gg 1$, therefore for a large initial dispersion we have,
\begin{equation}
\overline{\Delta}_F(\alpha,\beta)\sim(\sin\alpha\cos\beta)^2,
\end{equation}
which gives a maximum entanglement for $\beta=\pm\pi/2$ for any $\alpha$ as showed in Fig. \ref{fig:6}.


\begin{thebibliography}{}
\bibitem{aharonov1993quantum} Aharonov, Y., Davidovich, L., Zagury, N.: Quantum random walks. Phys. Rev. A \textbf{48}, 1687 (1993).
\bibitem{kempe2003quantum} Kempe, J.: Quantum random walks: an introductory overview. Contemporary Physics \textbf{44}, 307–327 (2003).
\bibitem{venegas2012quantum} Venegas-Andraca, S.E.: Quantum walks: a comprehensive review. Quantum Inf Process \textbf{11}, 1015–1106 (2012).
\bibitem{shenvi2003quantum} Shenvi, N., Kempe, J., Whaley, K.B.: Quantum random-walk search algorithm. Phys. Rev. A \textbf{67}, 052307 (2003).
\bibitem{engel2007evidence} Engel, G.S., Calhoun, T.R., Read, E.L., Ahn, T.K., Man{\v{c}}cal, T., Cheng, Y.C., Blankenship, R.E., Fleming, G.R.: Evidence for wavelike energy transfer through quantum coherence in photosynthetic systems. Nature \textbf{446}, 782–786 (2007).
\bibitem{childs2009universal} Childs, A.M.: Universal computation by quantum walk. Phys. Rev. Lett. \textbf{102}, 180501 (2009).
\bibitem{lovett2010universal} Lovett, N.B., Cooper, S., Everitt, M., Trevers, M., Kendon, V.: Universal quantum computation using the discrete-time quantum walk. Phys. Rev. A \textbf{81}, 042330 (2010).
\bibitem{chandrashekar2013two} Chandrashekar, C.: Two-component dirac-like hamiltonian for generating quantum walk on one-, two- and three-dimensional lattices. Scientific Reports \textbf{3}, 2829 (2013).
\bibitem{wang2013physical} Wang, J., Manouchehri, K.: Physical implementation of quantum walks. Springer (2013).
\bibitem{abal2006quantum} Abal, G., Siri, R., Romanelli, A., Donangelo, R.: Quantum walk on the line: Entanglement and nonlocal initial conditions. Phys. Rev. \textbf{73}, 042302 (2006).
\bibitem{abal2008erratum} Abal, G., Siri, R., Romanelli, A., Donangelo, R.: Erratum: Quantum walk on the line: Entanglement and non-local initial conditions [phys. rev. a 73, 042302 (2006)]. Phys. Rev. A 73, 069905 (2006).
\bibitem{salimi2012asymptotic} Salimi, S., Yosefjani, R.: Asymptotic entanglement in 1d quantum walks with a time-dependent coined. International Journal of Modern Physics B \textbf{26}, 1250112 (2012).
\bibitem{eryiugit2014time} Eryi{\u{g}}it, R., G{\"u}nd{\"u}{\c{c}}, S.: Time exponents of asymptotic entanglement of discrete quantum walk in one dimension. International Journal of Quantum Information \textbf{12}, 1450036 (2014).
\bibitem{alles2012maximal} All{\'e}s, B., G{\"u}nd{\"u}{\c{c}}, S., G{\"u}nd{\"u}{\c{c}}, Y.: Maximal entanglement from quantum random walks. Quantum Inf Process \textbf{11}, 211–227 (2012).
\bibitem{vieira2013dynamically} Vieira, R., Amorim, E.P.M., Rigolin, G.: Dynamically disordered quantum walk as a
maximal entanglement generator. Phys. Rev. Lett. \textbf{111}, 180503 (2013).
\bibitem{vieira2014entangling} Vieira, R., Amorim, E.P.M., Rigolin, G.: Entangling power of disordered quantum walks. Phys. Rev. A \textbf{89}, 042307 (2014).
\bibitem{valcarcel2010tailoring} de Valc\'arcel, G.J., Rold\'an, E., Romanelli, A.: Tailoring discrete quantum walk dynamics via extended initial conditions. New Journal of Physics \textbf{12}, 123022 (2010).
\bibitem{zhang2016creating} Zhang, W.W., Goyal, S.K., Gao, F., Sanders, B.C., Simon, C.: Creating cat states in one-dimensional quantum walks using delocalized initial states. New Journal of Physics \textbf{18}, 093025 (2016).
\bibitem{romanelli2010distribution} Romanelli, A.: Distribution of chirality in the quantum walk: Markov process and entanglement. Phys. Rev. A \textbf{81}, 062349 (2010).
\bibitem{romanelli2012thermodynamic} Romanelli, A.: Thermodynamic behavior of the quantum walk. Phys. Rev. A \textbf{85}, 012319 (2012).
\bibitem{romanelli2014entanglement} Romanelli, A., Segundo, G.: The entanglement temperature of the generalized quantum walk. Physica A \textbf{393}, 646–654 (2014).
\bibitem{bennett1996concentrating} Bennett, C.H., Bernstein, H.J., Popescu, S., Schumacher, B.: Concentrating partial entanglement by local operations. Phys. Rev. A \textbf{53}, 2046 (1996).
\bibitem{nayak2001one} Ambainis, A., Bach, E., Nayak, A., Vishwanath, A., Watrous, J.: One-dimensional quantum walks. In: Proceedings of the thirty-third annual ACM symposium on Theory of computing, pp. 37–49. ACM (2001).
\bibitem{ide2001entanglement} Ide, Y., Konno, N., Machida, T.: Entanglement for discrete-time quantum walks on the line. Quantum Information and Computation, Vol. 11, No.9\&10, pp.855-866 (2011).
\bibitem{nielsen2010quantum} Nielsen, M.A., Chuang, I.L.: Quantum computation and quantum information. Cambridge university press (2010).
\bibitem{abramowitz1964handbook} Abramowitz, M., Stegun, I.A.: Handbook of mathematical functions: with formulas, graphs, and mathematical tables, vol. 55. Courier Corporation (1964).
\end{thebibliography}
\end{document}